\documentclass[12pt]{iopart}
\pdfoutput=1
\expandafter\let\csname equation*\endcsname\relax
\expandafter\let\csname endequation*\endcsname\relax
\usepackage[numbers,sort&compress]{natbib}
\usepackage{amsmath,amssymb,eucal,graphicx,subfigure}
\usepackage{eso-pic,calc}
\usepackage{graphicx}
\usepackage{epsfig}
\usepackage{bm}
\usepackage{color}
\usepackage{wasysym}

\usepackage[colorlinks=True]{hyperref}  
\hypersetup{
 	colorlinks=true,  
 	linkcolor=blue,  
 	citecolor=blue, 
 	filecolor=magenta,   
 	urlcolor=blue         
 }

\def\beqr{\begin{eqnarray}}
	\def\eqnr{\end{eqnarray}}
\def\beq{\begin{equation}}
	\def\bc{\begin{center}}
		\def\ec{\end{center}}
	\def\eqn{\end{equation}}

\def\etl{$et~al.$~}

\begin{document}
\title{Avalanche activity noises in sandpile models}

\author{Rahul Chhimpa}
\address{Department of Physics, Institute of Science,  Banaras Hindu University, Varanasi 221 005, India}

\author{Avinash Chand Yadav\footnote{jnu.avinash@gmail.com}}
\address{Department of Physics, Institute of Science,  Banaras Hindu University, Varanasi 221 005, India}

\begin{abstract}
{We consider the Bak-Tang-Wiesenfeld (BTW) and the Manna sandpile models of self-organized criticality. In the models, previous studies revealed a signature of long-range temporal correlations in the avalanche activity. We examine the power spectra of the noises with different system sizes and find that the power spectrum for a finite-size system exhibits three distinct frequency regimes: (i) a frequency-independent behavior below a lower cutoff frequency, (ii) a hump-type behavior in the intermediate-frequency regime, and (iii) a power-law scaling $1/f^{\alpha}$ in the high-frequency regime. The power scales with the system size in all regimes but with different exponents. Also, the lower cutoff and peak frequencies decay in a power-law manner with the system size. We apply finite-size scaling and obtain data collapse for the power spectra, corroborating the estimation of the scaling exponents. Our studies reveal subtle scaling features for the temporal correlation within the avalanches.}
\end{abstract}

\maketitle

\section{Introduction}
A class of natural systems exhibits scaling features in space-time correlations. Self-organized criticality (SOC) offers one of the plausible routes explaining the emergent scaling behavior~\cite{PhysRevLett.59.381, PhysRevA.38.364, Bak_1996, Jensen_1998, Pruessner_2012, MARKOVIC201441}. The phenomena of interest appear in diverse contexts ranging from seismic activity~\cite{bak_1989, Sornette_1989, PhysRevLett.68.1244, ida_2024} and species evolution~\cite{PhysRevLett.71.4083, PhysRevE.108.044109} to neuronal activity~\cite{Beggs_2003, Bornholdt_2003, Chialvo_2010} and economic fluctuations~\cite{scheinkman_1994, tebaldi_2021}. One can demonstrate the key idea of SOC with a sandpile. A slow addition of sand grains can cause instability in the system. Responding to the instability, the system attains a stable configuration through an instantaneous cascade event, an avalanche. During the propagation of avalanches, the sand grains can leave the system (dissipation). The avalanche sizes can have all possible length scales and satisfy a power-law decaying probability distribution~\cite{pradhan_2021, denis_2024}.

Bak~\etl proposed a paradigm of SOC with cellular automata dynamics, popularly known as the BTW sandpile model~\cite{PhysRevLett.59.381}. The model consists of a two-dimensional (2D) medium as a square lattice, where each site can have a certain number of sand grains. At a site, the height of sand grains in a stable configuration remains $0\leq h_i < h_c$ below a threshold height $h_c = 4$. An addition of a sand grain at a randomly chosen site represents a slow external drive. The driving can yield an unstable configuration where site(s) can have a height equal to or larger than $h_c$. The system relaxes through a cascade of toppling events. Removing four sand grains from the unstable site and transferring one grain to each nearest neighbor constitutes a basic toppling rule. The toppling occurs in parallel until all sites become stable.

Each cascade event forms an avalanche whose size (the total toppled sites) remains a random variable. The sand grains do not fall on the system during the toppling process to avoid interacting avalanches. The system evolves following the same dynamics iteratively, starting from a random or flat initial configuration. After a sufficiently long time, the system reaches a critical state. In the stationary state, the avalanche sizes follow a power-law $P(s) \sim s^{-\tau_s}$. The model is simple, yet the exact value of the critical exponent $\tau_s$ remains unknown. Strikingly, the avalanche size distribution does not follow finite-size scaling (FSS)~\cite{PhysRevE.58.R2677, PhysRevLett.83.3952}. There has been a debate about the numerically estimated values of the exponent. A recent study reveals numerically $\tau_s \approx 1$ along with a system size-dependent behavior for the probability~\cite{PhysRevE.106.014148}. So far, analytical studies remain limited~\cite{PhysRevLett.63.1659, DHAR19994, DHAR200629}. However, several studies examined numerical variants of the BTW sandpile model. The typical variants include continuous energy sandpile~\cite{PhysRevE.56.1590}, directed sandpile~\cite{PhysRevA.39.6524, Tsuchiya_1999, PhysRevE.93.042107}, stochastic sandpile~\cite{Manna_1991, DHAR199969}, and fixed energy sandpile~\cite{PhysRevLett.81.5676, PhysRevE.57.5095, PhysRevLett.104.145703}.

\begin{figure}[t]
	\centering
	\scalebox{1.0}{\includegraphics{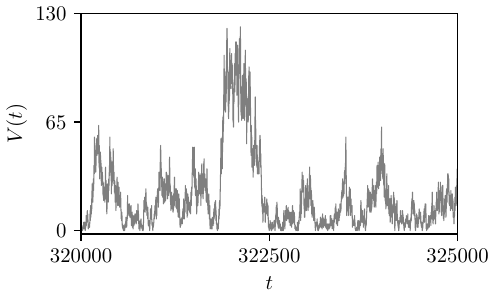}}
	\caption{A portion of the avalanche activity signal $V(t)$ in the BTW model with system size $N=2^{14}$. The noise $V(t)$ denotes the total number of toppled sites as a function of the fast time $t$. When an avalanche event finishes, a new sand grain falls on the system at a randomly chosen site. $V(t) = 0$ if no toppling occurs.}
	\label{fig_btw_0}
\end{figure}

Another interesting aspect in the BTW model is the avalanche activity noise $V(t)$, the number of toppled sites as a function of the fast time $t$ (cf figure~\ref{fig_btw_0}). The key relevance of this signal is to reveal temporal correlation within the avalanches. BTW numerically examined the temporal correlation of the process, and they found the power spectral density to follow  $1/f^{\alpha}$-type scaling behavior with the spectral exponent $\alpha \approx 1$~\cite{PhysRevLett.59.381}. However, later studies recognized the value of $\alpha$ to be incorrect. For small system sizes, the spectral exponent is $\alpha \approx 2$, implying Lorentzian form~\cite{PhysRevB.40.7425, Kertesz_1990}. However, Laurson~\etl performed large-scale simulations, indicating the spectral exponent is significantly smaller than 2~\cite{Laurson_2005}. Interestingly, they argued that the spectral exponent equals a critical exponent relating to a scaling between the average avalanche size and duration. Similarly, the Manna sandpile (a stochastic variant of the BTW model) exhibits qualitatively similar features in different dimensions~\cite{Laurson_2005}. The avalanche activity signal remains extensively studied in other contexts. For example, Barkhausen noise in ferromagnets shows $1/f^{\alpha}$ noise with a nontrivial spectral exponent in the high-frequency regime~\cite{PhysRevB.46.10822, PhysRevB.62.11699, PhysRevB.76.024406, stefano_2011, bohn_2018, bosi_2019}.

It is worth emphasizing that recent works have revealed that the $1/f^{\alpha}$ noises in several SOC systems can also exhibit a nontrivial system size scaling~\cite{PhysRevE.104.064132, PhysRevE.108.044109, PhysRevE.109.054130, PhysRevE.110.034130, PhysRevE.111.024108}. Motivated by this, we reexamine numerically the avalanche activity noises in two SOC systems: the BTW and the Manna models. Our extensive simulation studies reveal subtle system size scaling not reported previously. The power spectrum for a finite-size system displays three distinct regimes. (i) A frequency-independent behavior below a lower cutoff frequency (inverse of the correlation time) emerges, describing the trivial uncorrelated activity. (ii) A hump-type feature occurs in the intermediate-frequency regime. Laurson \etl mention the hump feature qualitatively but lack quantitative characterization. (iii) In the high-frequency regime, the power decreases in a power-law manner $\sim 1/f^{\alpha}$. When the system size increases, the power interestingly grows with the system size in a power-law in the entire frequency regime. A system size scaling also appears for the lower cutoff frequency $f_0 \sim N^{-z/{\rm D}}$ and the peak frequency $f_p \sim N^{-z_p/{\rm D}}$ corresponding to the maximum power value. One can also apply FSS to obtain data collapse for the power spectra of the two nontrivial frequency regimes to validate the estimation of the scaling exponents~\cite{Singh_2024}.

The plan of the paper is as follows: Section~\ref{sec_2} presents the definition of the sandpile models and their contrasting features. In section~\ref{sec_3}, we report our numerical results obtained from the Monte Carlo simulation of the models. We show the system size scaling features for the power spectra of the avalanche activity fluctuations. Finally, the paper concludes with future opportunities in section~\ref{sec_4}.

\section{Models}{\label{sec_2}}
In this section, we recall sandpile models. Specifically, we study the BTW and the Manna models. The original sandpile remains extensively studied and popularly known as the BTW model. Consider a square lattice as a medium with $N$ sites. At each site, assign an integer $h_i$ representing the height of sand grains. The sites are stable if the height lies such that $0\leq h_i < h_c$ with the threshold height $h_c = 4$. The dynamics include two steps: (i) Perturbation: An addition of a sand grain on a randomly chosen site represents a slow external drive
\begin{equation}
	h_i \to h_i+1. \nonumber
\end{equation}
As a result of the drive, the system's configuration may become unstable, or the site may attain a height equal to or greater than the threshold $h_i\geq h_c$. (ii) Relaxation: The system relaxes by a rearrangement of the heights. The unstable site reduces its height by $h_c$ and transfers the sand grains to the nearest neighbor sites, one grain to each as
\begin{equation}
	h_i \to h_i-h_c~~~ {\rm and}~~~h_j \to h_j +1, \nonumber
\end{equation} 
where $j$ denotes the nearest neighbors~\cite{PhysRevLett.59.381}. Consequently, the neighbor site(s) may become unstable, and the same toppling rule applies in parallel and recursively till all sites become stable. The boundaries are open, allowing dissipation of sand grains. The relaxation rule is deterministic and locally conservative.

Initially, one can consider a random configuration for the height of the sand grains. After evolving for a long time, the system moves from a transient to a stationary regime. Some configurations that appear in the transient but don't occur in the stationary regime are termed forbidden configurations. In the stationary regimes, the configurations are recurrent~\cite{PhysRevLett.64.1613, PhysRevE.108.014108, manna2025}. To identify if a state is transient or recurrent, one can use the burning algorithm~\cite{MAJUMDAR1992129}.

The number of toppled sites $V(t)$ as a function of the fast time $t$ denotes the avalanche activity noise~\cite{Laurson_2005}. The total toppling per avalanche event (the distinctly toppled sites) denotes the avalanche size (area). The avalanche size (area) follows a power-law decaying behavior with a cutoff size limited by the system size. An addition of a new sand grain happens only when the avalanche event ends to avoid overlapping of multiple avalanches. Thus, a separation of time scales emerges with the slow drive and instantaneous dissipation.

In the Manna model, the threshold height remains $h_c = 2$, independent of the dimension~\cite{Manna_1991, DHAR199969}. The toppling rule becomes stochastic but remains locally conservative. If a site is unstable, two sand grains are removed and transferred to two nearest neighbors, where each potential nearest neighbor is chosen randomly and independently. The two models belong to different universality classes. The avalanche size distribution follows FSS for the Manna model but breaks for the BTW model.

\begin{figure}[t]
	\centering
	\scalebox{1.0}{\includegraphics{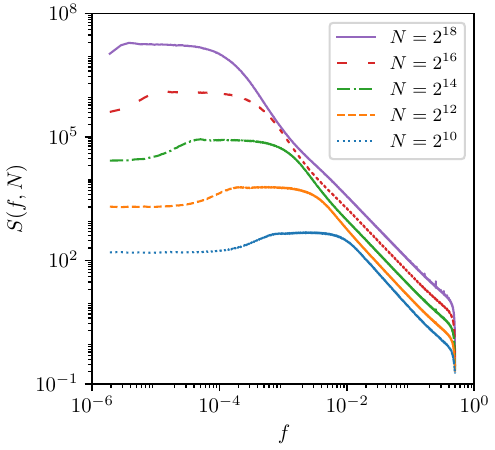}}
	\caption{The power spectra $S(f, N)$ for the signals $V(t)$ in the BTW model for different system sizes $N = 2^{10}, 2^{12}, 2^{14}, 2^{16}$, and $2^{18}$. After discarding transients, we record the signal for time up to $2^{20}$. Each curve represents an average over $10^4$ independent realizations of the noise.} 
	\label{fig_btw_1}
\end{figure}

\begin{figure}[t]
	\centering
	\scalebox{1.0}{\includegraphics{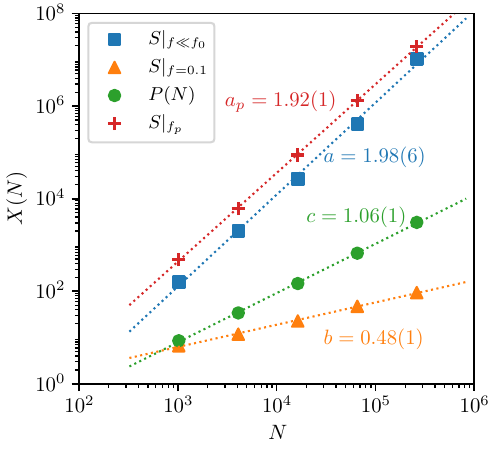}}
	\caption{The system size scaling for different quantities along with the best-fit (dotted lines). The symbols denote the power $S|_{f\ll f_0}$ (below the lower cutoff frequency), $S|_{f=0.1}$ (in the high-frequency regime), the total power $P(N)$, and $S|_{f_p}$ (the peak power).}
	\label{fig_btw_2}
\end{figure}

\begin{figure}[t]
	\centering
	\scalebox{1.0}{\includegraphics{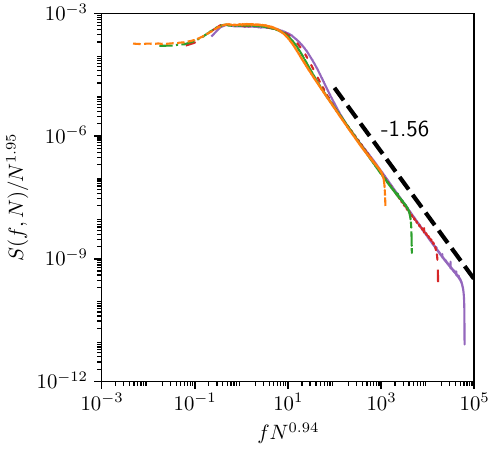}}
	\caption{The data collapse of the power spectra shown in figure~\ref{fig_btw_1}. The dashed line guides the slope.}
	\label{fig_btw_3}
\end{figure}

\section{Results}{\label{sec_3}}
Figure~\ref{fig_btw_0} shows a typical plot of the avalanche activity signal $V(t)$ as a function of the fast time $t$ in the BTW model. In this paper, we aim to understand the underlying temporal correlation. We compute the power spectral density (PSD), the Fourier transformation of the two-time autocorrelation function of $V(t)$, as the noise is a weakly stationary process. To compute the PSD, we need to determine $\tilde{V}(f, N)$, the Fourier transformation of the signal. We implement the standard fast Fourier transform algorithm on the signal with length $T$. Then, we take the square of the modulus of $\tilde{V}(f, N)$. The power spectrum computed from a single realization remains noisy. We perform averaging over $M$ independent realizations to smooth it. We can express the PSD as a function of frequency and the system size as  
\begin{equation}
	S(f, N) = \lim_{T\to \infty}\frac{1}{T} \langle |\tilde{V}(f, N)|^2\rangle,\nonumber
	\label{eq_1}
\end{equation}
where $\langle \cdot \rangle$ the angular bracket denotes the average.

\begin{figure}[t]
	\centering
	\scalebox{1.0}{\includegraphics{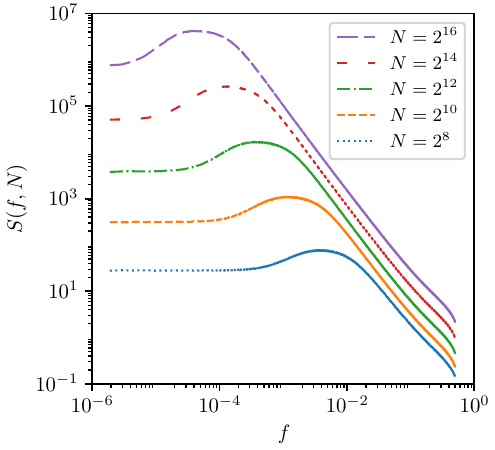}}
	\caption{Same as figure~\ref{fig_btw_1} but for the Manna model in 2D. }
	\label{fig_manna_2d_1}
\end{figure}

\begin{figure}[t]
	\centering
	\scalebox{1.0}{\includegraphics{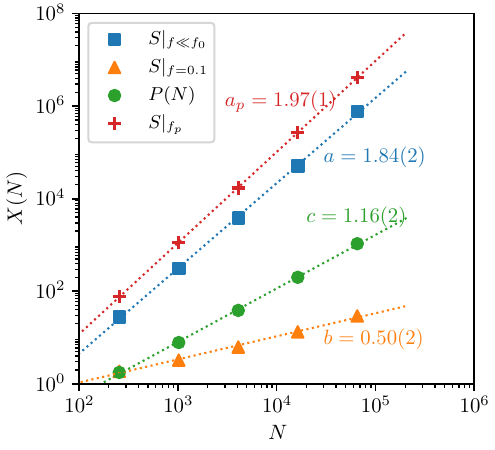}}
	\caption{Same as figure~\ref{fig_btw_2} for the 2D Manna model. }
	\label{fig_manna_2d_2}
\end{figure}

\begin{figure}[t]
	\centering
	\scalebox{1.0}{\includegraphics{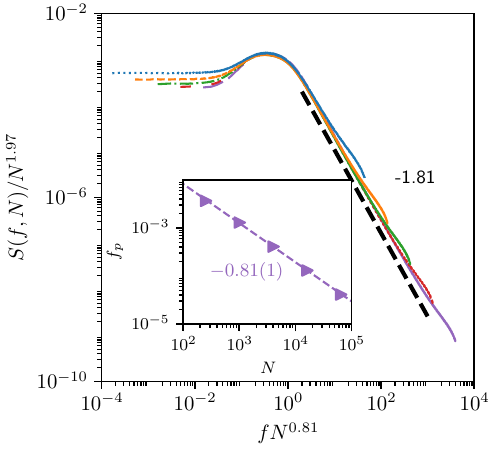}}
	\caption{Same as figure~\ref{fig_btw_3} for the 2D Manna model. The inset shows system size scaling of the peak frequency.}
	\label{fig_manna_2d_3}
\end{figure}

\begin{figure}[t]
	\centering
	\scalebox{1.0}{\includegraphics{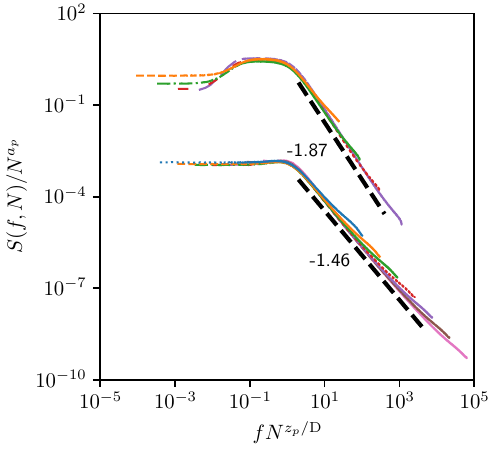}}
	\caption{The power spectra scaling functions for the Manna model in 1D (lower) and 3D (upper curves). For the 3D Manna model, the curves are shifted vertically by a constant factor for clarity. The system sizes are $N = 2^5, 2^6, 2^7, 2^8, 2^9, 2^{10},$ and $2^{11}$ in 1D and $N = 2^9, 2^{12}, 2^{15},$ and $2^{18}$ in 3D. In the intermediate frequency regime, the hump-type behavior in the 1D Manna model is not profound.}
	\label{fig_manna_1_3d_3}
\end{figure}

As shown in figure~\ref{fig_btw_1}, the power spectrum on a double logarithmic scale exhibits three distinct frequency regimes in the BTW model. One can note the following features for a finite system size: (i) The frequency-independent behavior below a lower cutoff frequency trivially displays uncorrelated activity. (ii) A hump-type behavior emerges in the intermediate frequency regime. (iii) A power-law decay of $1/f^{\alpha}$-type behavior appears in the high-frequency regime. The power also diverges with the system size following a scaling behavior with a critical exponent that differs for different frequency regimes. Moreover, the lower cutoff and the peak frequencies decay in a power-law with the system size. We can summarize the scaling features mathematically as  
\begin{equation}
	S(f, N) \sim \begin{cases} N^a, ~~~~~~~~~~~~ f\ll N^{-z/{\rm D}}, \\ N^{a_p},~~~~~~~~~~~ f_p = N^{-z_p/{\rm D}}, \\ N^b/f^{\alpha},~~~~~~N^{-z_p/{\rm D}} \ll f \ll 1/2.\end{cases} 
	\label{eq_2}
\end{equation}
In addition, the total power also diverges as $P(N) \sim \int df S(f, N) \sim N^{c}$.
Figure~\ref{fig_btw_2} shows the system size scaling of the power values in different frequency regimes, including the total power.  We obtain the critical exponents using best-fit.

We apply the FSS to obtain data collapse for the power spectra, validating the estimation of the critical exponents. We scale the frequency by the peak frequency as $fN^{z_p/{\rm D}}$ and the power as $S(f, N)/N^{a_p}$ to obtain the data collapse of the two nontrivial frequency regimes. However, the scaled power in the low-frequency regime varies marginally as $\sim N^{a-a_p}$. In the high-frequency regime, the scaling function shows size-independent behavior. Equation~(\ref{eq_2}) suggests $S(f, N)/N^{a_p} \sim N^b/f^{\alpha}N^{a_p} \sim N^{\alpha z_p/{D}}/N^{a_p-b} (fN^{z_p/{\rm D}})^{\alpha}$, implying 
\begin{equation}
	\alpha = {\rm D}(a_p-b)/z_p.
	\label{eq_3}
\end{equation}

Figure~\ref{fig_btw_3} displays the data collapse for the power spectra of the signals $V(t)$ in the BTW model. In this case, the hump-type behavior is not well-rounded but a bit flat. The error bar for the exponent $a$ is significantly large due to the finite signal length. The critical exponents used to obtain the scaling function differ slightly from the estimated exponents from the best-fit.

Following a similar analysis, we examine the avalanche signals in the Manna model in dimensions ranging from 1D to 3D. Qualitatively, we get similar results (cf figures~\ref{fig_manna_2d_1}-\ref{fig_manna_2d_3} for the 2D Manna model). In the 2D Manna model, the hump-type feature is well-rounded about a peak power. Particularly, the hump-type behavior is no longer profound in 1D, and it remains flat in the 3D model, similar to the BTW case (cf figure~\ref{fig_manna_1_3d_3}). The critical exponents tabulated in Table~\ref{tab1} provide a comparison of the two models. We emphasize that the spectral exponent estimated using equation~(\ref{eq_3}) consistently agrees with the previous studies~\cite{Laurson_2005}.

\begin{table}[]
	\centering
	\begin{tabular}{ccccccc}
		\hline 
		
		Model    & $\alpha$ &  $z_p$& $ z$ & $a_p$ & $a$ & $b$ \\		
		\hline
		\hline
		BTW      & 1.56(4) & 1.88(2)    & 1.8(1)   & 1.92(1) & 1.98(6) & 0.48(1) \\
		1D Manna & 1.46(7) & - & 1.54(5) & -   & 2.87(3) & 0.62(1)     \\
		2D Manna &1.81(6) & 1.62(2) & 1.36(8) & 1.97(1) & 1.84(2)   & 0.50(2)     \\
		3D Manna &1.9(2) & 1.9(2) & 1.3(2) & 1.51(3) &1.32(4) & 0.34(2)     \\
		\hline
	\end{tabular}
	\caption{The critical exponents as obtained numerically characterize the avalanche activity noise power spectra in the BTW and the Manna models. We use equation~(\ref{eq_3}) to compute $\alpha$. For the BTW model, we estimate the spectral exponent using the values of $a_p$ as used in the data collapse (cf figure~\ref{fig_btw_3}).}
	\label{tab1}
\end{table}

\section{Conclusion}{\label{sec_4}}
We consider the BTW and Manna sandpile models, which demonstrate self-organized criticality. One of the quantities of interest as a function of the fast time is the avalanche activity noise that can offer insight into the temporal correlation within the avalanche events. Previous studies revealed that the process exhibits a scaling feature in the power spectrum as $1/f^{\alpha}$ with the spectral exponent being significantly smaller than 2, which signifies the presence of long-range temporal correlations. Notice that the spectral exponent is equal to an exponent describing the scaling between average avalanche size and duration.

Our extensive simulation studies reveal intriguing system size scaling features of the power spectra of the noises. 
When the system size increases, the power interestingly grows with the system size in a power-law in the entire frequency regime, where the scaling exponent takes a distinct value in different frequency regimes. In the high-frequency regime, the power scales as $\sim N^{1/{\rm D}}$ for $1<{\rm D}\leq 3$. Three distinct frequency regimes emerge in the power spectrum for a finite system. (i) A frequency-independent behavior in the low-frequency regime describes the trivial uncorrelated activity $\sim 1/f^0$. The underlying correlation time diverges with the finite system size as $\sim N^{z/{\rm D}}$. (ii) A hump occurs in the intermediate frequency regime. The average time corresponding to the peak power also diverges as $\sim N^{z_p/{\rm D}}$. The peak power implies the presence of a regular temporal event, referring to the external drive that occurs after an average time. Numerically, the two dynamical exponents differ marginally.  (iii) The power-law scaling of $1/f^{\alpha}$-type arises in the high-frequency regime. The process shows a long-range temporal correlation in the short-time scale.   It is also possible to apply finite-size scaling to obtain data collapse for the power spectra in the nontrivial frequency regimes, validating the estimation of the scaling exponents. The scaled power in the low-frequency regime varies as $N^{a-a_p}$, where the exponent $a-a_p$ remains significantly small. We infer that one may attribute the emergence of different regimes to a competition between the finite observation time and the correlation time or the average time after which the external drive occurs.

\section*{ACKNOWLEDGMENTS}
RC acknowledges UGC, India, for financial support through a Senior Research Fellowship. We greatly acknowledge the PARAM Shivay, the Supercomputing Centre, IIT BHU, Varanasi.

\bibliography{s1sources}
\bibliographystyle{myrev}
\end{document}